# Observing oscillations using the non-magnetic and magnetic spin Hall effect in a non-collinear antiferromagnet

Monalisa Cavalcante[1,2], José Laurentino[1,2], Carlos Eduardo[1,3], Kevin Mesel[1,4], and José Holanda[1,2,3,4,5,*]

[1]Group of Optoelectronics and Spintronics, Universidade Federal Rural de Pernambuco, 54518-430, Cabo de Santo Agostinho, Pernambuco, Brazil
[2]Programa de Pós-Graduação em Engenharia Física, Universidade Federal Rural de Pernambuco, 54518-430, Cabo de Santo Agostinho, Pernambuco, Brazil
[3]Programa de Pós-Graduação em Física Aplicada, Universidade Federal Rural de Pernambuco, 52171-900, Recife, Pernambuco, Brazil
[4]Unidade Acadêmica do Cabo de Santo Agostinho, Universidade Federal Rural de Pernambuco, 54518-430, Cabo de Santo Agostinho, Pernambuco, Brazil
[5]Programa de Pós Graduação em Tecnologias Energéticas e Nucleares (Proten), Universidade Federal de Pernambuco, Recife, 50740-545, PE, Brazil

## Abstract

In this study, it was observed that, in non-collinear $IrMn_3$, both the non-magnetic spin Hall effect (NMSHE) and the magnetic spin Hall effect (MSHE) exhibit oscillations as the triangularly ordered magnetic moments of $IrMn_3$ evolve with temperature. These oscillations are strongly influenced by the magnetic properties at the interface between $IrMn_3$ and the adjacent material. Permalloy (Py), a ferromagnet, and platinum (Pt), a paramagnet, were employed as adjacent layers. Such experimental analyses demonstrate the pioneering nature of this phenomenon, which is validated through magnetoresistance measurements performed under different experimental conditions.

*Author corresponding: joseholanda.silvajunior@ufrpe.br

## I. INTRODUCTION

Symmetry-breaking mechanisms can be used to characterize different phases of matter. In antiferromagnetic (AFM) materials, the rotational symmetry present in the paramagnetic state is broken by the Néel vector. The so-called easy axes - preferred crystallographic directions of the magnetic moments - are determined by the relevant order parameters, namely the Néel vector and spin-orbit coupling. These directions correspond to minima of the magnetocrystalline anisotropy energy (MAE). Certain AFMs, such as $Mn_2Au$, are well approximated as isotropic Heisenberg magnets and exhibit an in-plane MAE on the order of 10 μeV per formula unit [1]. In contrast, a particularly interesting non-collinear AFM, $IrMn_3$, exhibits a giant MAE of approximately 10 meV per formula unit [2]. An important feature of $IrMn_3$ is its large spin Hall angle, reaching values up to ~0.35 along the [001] direction [3]. The structural quality of $IrMn_3$ has been extensively characterized using X-ray diffraction (XRD) [4], neutron scattering [5], and theoretical calculations [6]. At the surface of thin antiferromagnetic films such as $IrMn_3$, local ferromagnetism arising from uncompensated spins can induce symmetry breaking in an adjacent material [7]. The strength of the interaction between $IrMn_3$ and the adjacent layer depends on the temperature at which the interface is formed. These factors critically influence the behavior of both the non-magnetic spin Hall effect (NMSHE) and the magnetic spin Hall effect (MSHE) [8].

The conventional spin Hall effect, also referred to as the non-magnetic spin Hall effect (NMSHE), is characterized by transverse spin currents generated in non-magnetic conductors under an applied electric field, leading to spin accumulation near the sample surfaces [9–14]. Several approaches have been explored to investigate phenomena associated with spin Hall effects [15,16], and non-magnetic materials have been extensively studied to elucidate their spin Hall properties. Current-induced spin polarization arising from spin–orbit interaction [9–17] has attracted significant attention due to its potential applications in spin-transfer torque devices [18]. In contrast, the magnetic spin Hall effect (MSHE) can be interpreted in terms of symmetry-defined linear response functions and originates from the reactive counterpart of the dissipative spin response [8]. In this paper, we report the observation of oscillations driven by both the NMSHE and MSHE at $IrMn_3$/Py and $IrMn_3$/Pt interfaces, where the magnetic order is governed by interface symmetry and, consequently, by the orientation of the magnetic moments.

## II. RESULTS

All $IrMn_3$ films were grown using a magnetron sputtering system with a base pressure of approximately 1 x $10^{-7}$ Torr, an argon working pressure of 3 mTorr, and a substrate temperature of 570 °C. A Ti capping layer (2 nm) was used to protect the surface properties of $IrMn_3$ and the adjacent materials. The Ti, Py, and Pt layers were deposited at room temperature. **Fig. 1(a)** illustrates the distribution of crystallographic planes in the $IrMn_3$ unit cell. In this structure, the Mn magnetic moments lie parallel to the {111} planes and are aligned along the ⟨112⟩ directions. Figure 1(b) shows the X-ray diffraction (XRD) pattern of an MgO/$IrMn_3$(10 nm)/Ti(2 nm) sample, revealing a [100] crystallographic orientation. XRD measurements were performed using a Philips diffractometer with CuK-α radiation ($\lambda$ = 1.54 Å). For the $IrMn_3$ cubic structure, the interplanar spacing $d_{hkl}$ of the (*hkl*) family is related to the lattice parameter *a* through $d_{hkl} = a/(h^2+k^2+l^2)^{1/2}$. The interplanar distance $d_{hkl}$ was obtained from Bragg's law, $\lambda = 2d\sin\theta$, where θ is the Bragg angle. From the (200) peak shown in the inset of **Fig. 1(b)**, a lattice parameter of $a$ = 0.377 nm was extracted, which is consistent with the value reported in Ref. [19]. The thickness of the $IrMn_3$ film was calibrated by fitting X-ray reflectivity data, as shown in **Fig. 1(c)**. **Figs. 1(d)** and **1(e)** present magnetic measurements performed on an MgO/$IrMn_3$(10 nm)/Py(6 nm)/Ti(2 nm) sample using a Quantum Design Superconducting Quantum Interference Device (SQUID). **Fig. 1(d)** shows the hysteresis loop measured at 300 K after a field-cooling process under an applied magnetic field of +70 kOe along the [001] crystallographic direction. **Fig. 1(e)** displays the derivative of the magnetization with respect to the applied magnetic field corresponding to the hysteresis loop in **Fig. 1(d)**. A small shift of the hysteresis loop is observed in **Figs. 1(d)** and **1(e)**, indicating the presence of exchange bias at room temperature. The exchange bias field is defined as $H_{EX} = -(H_1 + H_2)/2$, while the coercivity is given by $H_C = (H_1 - H_2)/2$, where $H_1$ and $H_2$ are the switching fields. The observation of exchange bias at 300 K implies that the blocking temperature is above room temperature, demonstrating that exchange bias can be stabilized at room temperature in the $IrMn_3$(10 nm)/Py bilayer [20].

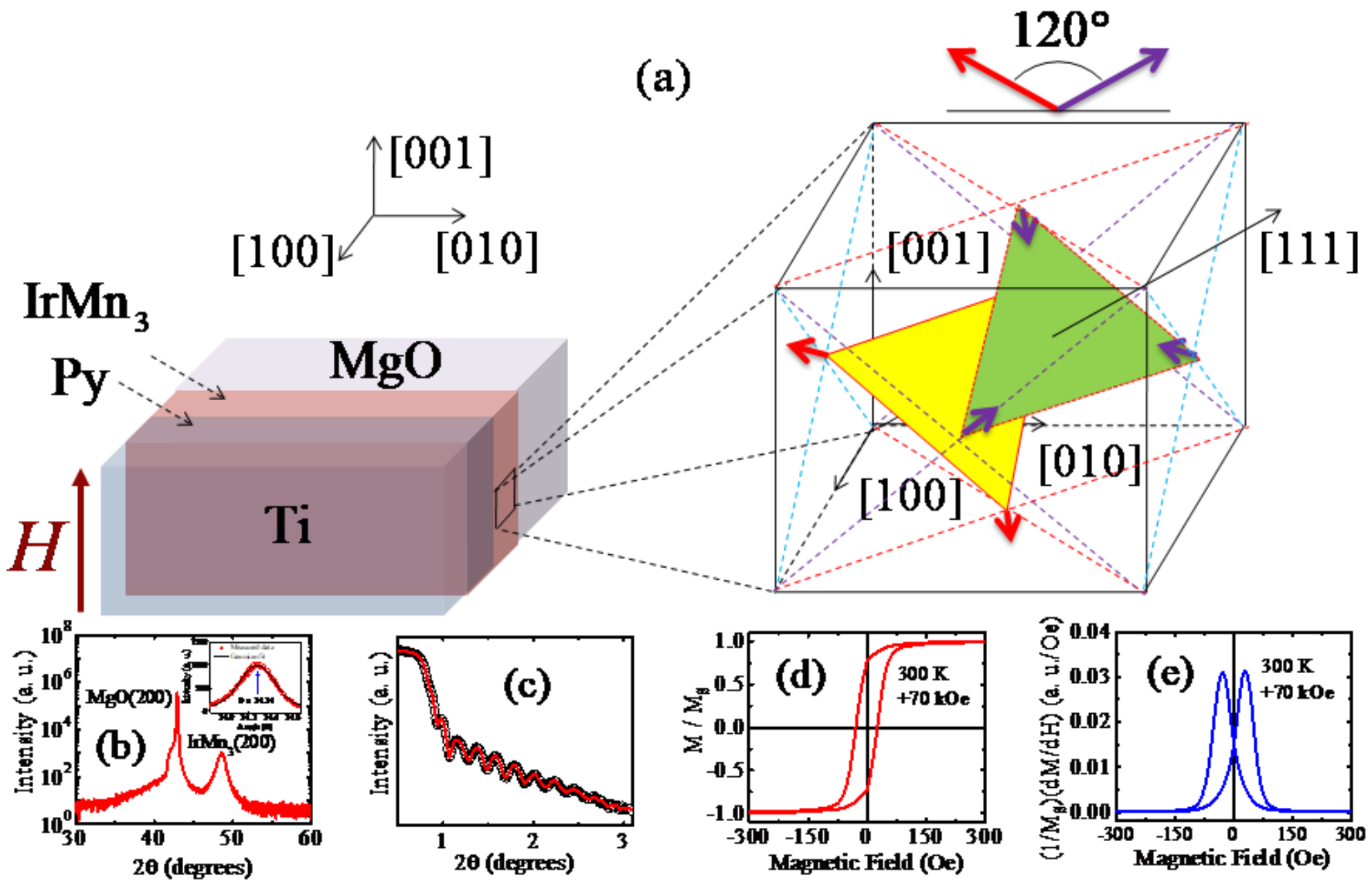

**Figure 1: (a)** Schematic illustration of the sample structures used in this work. The unit-cell arrangement of the antiferromagnetic $IrMn_3$ is also shown, highlighting its non-collinear spin configuration with a 120° angle between neighboring spins; **(b)** X-ray diffraction (XRD) pattern of a 10-nm-thick (100)-oriented $IrMn_3$ film capped with a 2-nm-thick Ti layer; **(c)** X-ray reflectivity (XRR) spectrum of a calibration $IrMn_3$ thin film (~20 nm). The solid red line represents the best fit used to determine the film thickness; **(d)** Magnetic hysteresis loop measured at 300 K after a field-cooling process under an applied magnetic field of +70 kOe, showing a small shift in the magnetic field; **(e)** Derivative of the magnetization with respect to the magnetic field obtained from the hysteresis loop shown in panel (d).

The exchange bias effect is generally attributed to uncompensated spins at the interface of an antiferromagnetic material, which are exchange-coupled to the magnetic moments of an adjacent ferromagnetic layer [21, 22]. In most thin-film systems, these uncompensated antiferromagnetic spins are strongly pinned along the direction of unidirectional anisotropy by antiferromagnetic domains [23, 24]. The difficulty associated with reorienting these AFM domains defines the blocking temperature ($T_B$); above this temperature, the energy barrier can be overcome, allowing the exchange bias direction to be reset by an applied magnetic field. A shift of the hysteresis loops due to the exchange bias effect is observed for both directions of the applied magnetic field: +70 kOe, as shown in **Fig. 2(a)**, and −70 kOe, as shown in **Fig. 2(b)**, with both measurements performed at 10 K. The insets of **Figs. 2(a)** and **2(b)** display the derivatives of the magnetization with respect to the applied magnetic field corresponding to the respective hysteresis loops. These insets clearly demonstrate that the exchange bias field follows the direction of the external magnetic field. The application of a sufficiently strong external magnetic field drives the ferromagnetic layer into saturation and simultaneously aligns the exchange-coupled uncompensated spins, leading to a uniform orientation of the antiferromagnetic domains [25, 26]. Upon cooling the heterostructure, the preferred antiferromagnetic domain orientation becomes fixed below $T_B$, and the exchange anisotropy is established along the direction of the applied field [23, 24]. The thermal stability of the resulting antiferromagnetic domain walls ultimately governs the magnitude and other characteristics of the exchange bias effect [27, 28].

The temperature dependence of the exchange bias field is shown in **Fig. 2(c)**. A monotonic decrease of the exchange bias field is observed with increasing temperature. At 380 K, the exchange bias effect vanishes, defining the blocking temperature ($T_B$) of the $IrMn_3$(10 nm)/Py bilayer [20]. The insets of **Fig. 2(c)** display the temperature dependence of the coercive fields $-H_1$ and $+H_2$, showing that at $T_B$ = 380 K the absolute values of the coercive fields become equal. At 10 K, the coercivity reaches its maximum value within the measured temperature range, ($H_C = -H_1+H_2 = -125$ Oe). As the sample is cooled, the coupling between the Ni magnetization in Py and the interfacial Mn moments becomes strongly pinned along a preferred direction determined by the dominant antiferromagnetic domain in the bulk of the $IrMn_3$ film [29, 30]. Below 100 K, the uncompensated Mn moments at the $IrMn_3$ interface with the adjacent material become antiferromagnetically ordered [31]. This transition explains the pronounced increase of the exchange bias field observed below 100 K, as shown in **Fig. 2(c)**. The temperature dependence of the uncompensated antiferromagnetic Mn moments at the $IrMn_3$ interface can be described by the power law $M(T) = M(T{=}10K)(1-T/T_N)\beta$, where $T_N$ is the Néel temperature of the interfacial Mn moments and $\beta$ is a critical exponent. For Heisenberg-type universality classes, reported values of $\beta$ include 9/2, 2, 3, 3/2, and 5/2; however, this regularity does not generally apply to Hamiltonians that deviate from a pure Heisenberg model. In the present case, using $\beta = 2$ [32] and $T_N$ = 100 K [31], the model provides a good fit to the experimental data in the temperature range close to 100 K.

The fit shown in **Fig. 2(d)** reveals antiferromagnetic interfacial behavior arising from the reorientation of the magnetic moments of Mn atoms at the $IrMn_3$/adjacent-material interface. Below the blocking temperature ($T_B$ = 380 K) for $IrMn_3$(10 nm)/Py), the thermal stability of the uncompensated Mn moments can be described by an exponential dependence, Exp[-E(T=10K)/$k_B$T] [33], where the maximum stability energy is E(T=10K) = 28.7 X$10^{-7}$ (erg/K) and $k_B$ = 1.38 X$10^{-16}$ (erg/K) is the Boltzmann constant. Considering the sample cross-sectional area of 2.25X$10^{-13}$ $cm^2$, an interfacial energy density of $E_{in}$ = 1.27x $10^{-2}$ erg/($Kcm^2$), is obtained. This value is comparable to the anisotropy energy density, defined as $E_k = M_S d_{FM} H_{EB}$, where $M_S$ is the saturation magnetization and $d_{FM}$ is the thickness of the ferromagnetic layer. Using $M_S$(T=10K) = 776 G **[34]**,$d_{FM}$ = 6 nm and $H_{EB}$ = 62.6 Oe we obtain $E_k$ = 2.9 x $10^{-2}$ erg/$cm^2$, which is of the same order of magnitude as reported in Ref. [35]. The temperature dependence of the magnetization was further analyzed by fitting the experimental data using a modified Bloch's law, $1 - m(T) = \alpha T^n$, where m(T) is the reduced magnetization and α, n are fitting parameters [36, 37]. The modified Bloch's law is applicable at elevated temperatures up to approximately 85% of the Curie temperature and allows indirect determination of transition-related parameters. Fitting the data in **Fig. 2(d)** over the temperature range 50 K < T < 390 K yields $\alpha = 0.62\ K^n$ and n = 0.06, providing a good description of the experimental behavior. This overall analysis explains why the exchange interaction at the $IrMn_3$/adjacent-material interface begins to increase significantly at temperatures around 100 K.

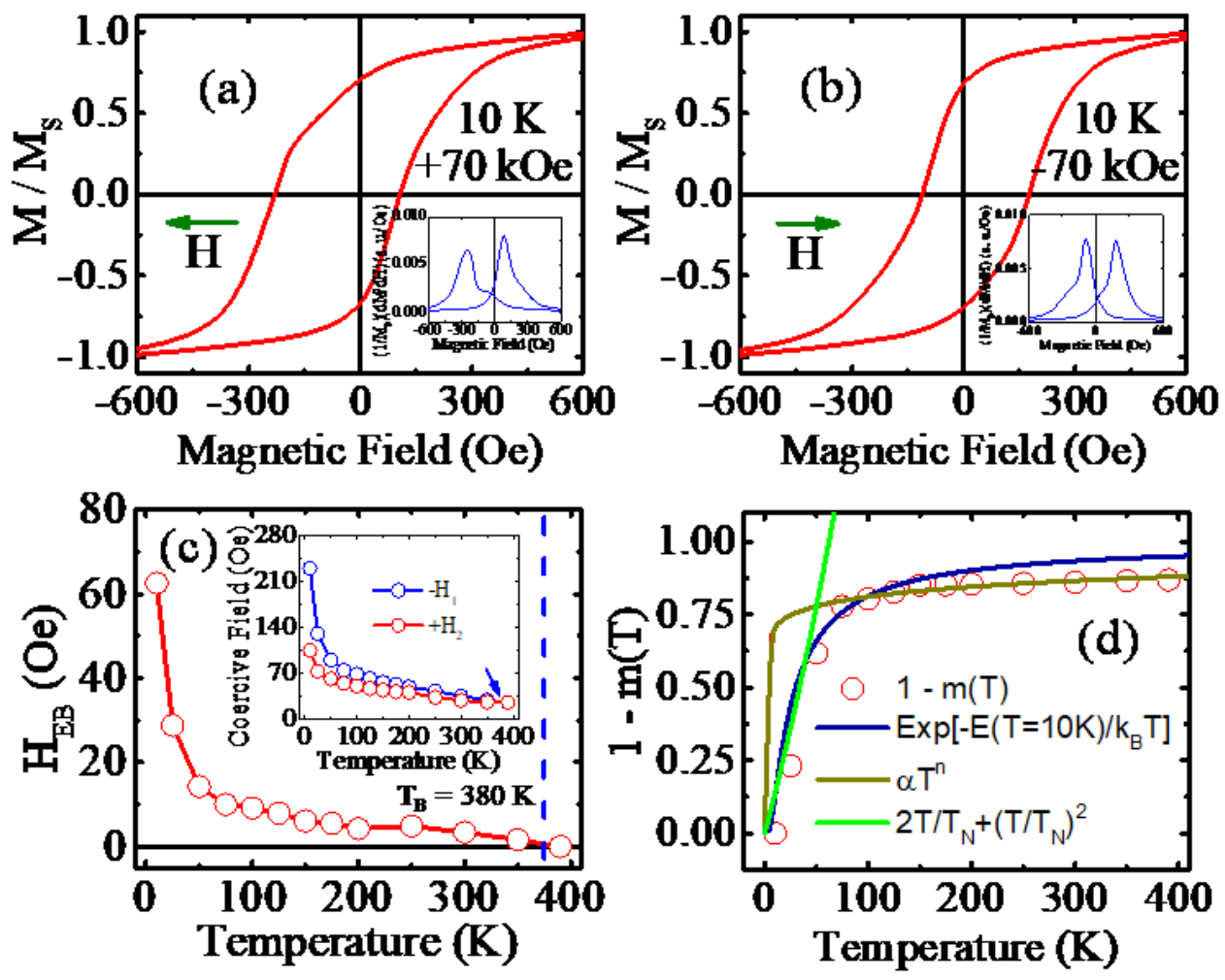

**Figure 2:** (a) and (b) Magnetic hysteresis loops measured at 10 K after a field-cooling process under applied magnetic fields of +70 kOe and −70 kOe, respectively, for an $IrMn_3$(10 nm)/Py(6 nm)/Ti(2 nm) sample. The insets show the derivatives of the magnetization with respect to the magnetic field. (c) Temperature dependence of the exchange bias field, revealing a blocking temperature of $T_B$ = 380 K. The inset shows the coercive fields as a function of temperature, indicating the locking temperature at which the coercivities for positive and negative fields become equal. (d) Temperature dependence of the reduced magnetization, defined as 1-m(T)=[M(T=10K)–M(T)]/M(T=10K). The function exponential exp[-E(T=10K)/$k_B$T] describes the thermal contribution and highlights the temperature range where thermal stability is most pronounced. The power-law behavior $\alpha T^n$ characterizes the regime dominated by ferromagnetic behavior in the temperature range [100K<T<390 K] where interfacial coupling instability effects become evident. The fit using 1-m (T) = $2T/T_N + (T/T_N)^2$ indicates the onset of a new exchange coupling, providing evidence of antiferromagnetic ordering of Mn atoms at the $IrMn_3$ surface.

From the results presented in **Fig. 2**, it can be observed that the interface between the antiferromagnet $IrMn_3$ and an adjacent material exhibits spintronic properties that differ from those of collinear and easy-plane antiferromagnets [38,39]. An efficient approach to investigating interfacial antiferromagnetic behavior is through the electrical detection of the surface magnetic state in AFMs [8]. In planar geometries, such access can be achieved via anisotropic magnetoresistance (AMR) measurements, as well as through non-magnetic and magnetic spin Hall effect measurements. The spatial distribution of surface magnetic moments in non-collinear antiferromagnets is of particular interest, since the distribution of uncompensated surface moments in these materials differs fundamentally from that in collinear antiferromagnets [38, 39]. The measurements shown in **Fig. 3** arise from three contributions: anisotropic magnetoresistance, the non-magnetic spin Hall effect, and the magnetic spin Hall effect. **Fig. 3(a)** illustrates the experimental configuration of the electrical contacts (contact 1 = $I_{DC}$ and contact 2 = R) used to measure the AMR, including contributions from both the magnetic and non-magnetic spin Hall effects, in MgO/$IrMn_3$(10 nm)/Py(6 nm)/Ti(2 nm) and MgO/$IrMn_3$(10 nm)/Pt(6 nm)/Ti(2 nm) samples, respectively. The crystallographic directions {[100], [010] and [001])} indicate the orientations of the applied magnetic field and DC current. **Fig. 3(b)** shows the resistance variation as a function of magnetic field measured with a DC current of 0.1 mA. The field shift indicated by the blue curve corresponds to the exchange bias field determined from transport measurements [22] at a temperature of 10 K. The inset of **Fig. 3(b)** confirms that the contacts shown in **Fig. 3(a)** exhibit ohmic behavior for the MgO/$IrMn_3$(10 nm)/Py(6 nm)/Ti(2 nm) sample.

As is well known, the critical current required to achieve steady-state stability in antiferromagnets can be lower than the corresponding values in ferromagnetic or paramagnetic systems. This reflects a fundamental difference between the current-induced dynamics of ferromagnetic (or paramagnetic) and antiferromagnetic materials under spin-polarized currents [40]. The resistance variation as a function of magnetic field, measured in the range 0–70 kOe and applied along different crystallographic directions, is shown in **Fig. 3(c)**. These measurements demonstrate a stable coupling at the $IrMn_3$/Py interface, with a uniform resistance variation throughout the entire magnetic-field range. This behavior arises from the strong exchange interaction discussed in **Fig. 2** and Ref. [35]. At 10 K, the exchange coupling is sufficiently strong that no self-oscillation effects are observed. Even at the maximum applied DC current (see the inset of **Fig. 2(b)**), no resistance changes attributable to the magnetic spin Hall effect (MSHE) are detected. A detailed angular dependence of the resistance, shown in **Fig. 3(d)**, reveals a smooth anisotropic magnetoresistance (AMR) behavior, further confirming that the MSHE is not strong enough to induce oscillations at this temperature. Similar phenomenological characteristics observed at the $IrMn_3$/Py interface are also present at the $IrMn_3$/Pt interface, particularly the stability of the exchange interaction. This indicates that the non-magnetic spin Hall effect (NMSHE) is likewise insufficient to destabilize the AMR and generate oscillations at this temperature, as shown in **Fig. 3(e)**. Comparing the resistance variations in **Figs. 3(e)** and **3(c)**, it is evident that the amplitude in **Fig. 3(e)** is significantly smaller. Specifically, the AMR amplitude at the $IrMn_3$/Pt interface, $\Delta R_{IrMn3/Pt} = R(\pi/2) - R(0) = 6.59\ \Omega$ is much smaller than that at the $IrMn_3$/Py interface, $\Delta R_{IrMn3/Py} = R(\pi/2) - R(0) = 0.0039\ \Omega$ corresponding to an amplitude ratio of approximately 1.69 x $10^5$ %. This large difference arises from the AMR contribution of permalloy [41]. In contrast, the AMR observed in **Fig. 3(e)** originates from the rotation of antiferromagnetic spins at the surface of $IrMn_3$ [42]. The inset of **Fig. 3(e)** confirms that both electrical contacts are ohmic. The AMR behavior shown in **Fig. 3(f)** is attributed to magnetic moments at the $IrMn_3$/Pt interface, where platinum behaves as a nearly ferromagnetic material due to the Stoner criterion [43]. These results indicate that the NMSHE contributes to the AMR signal but does not induce instability at this temperature. In this configuration, both the platinum and the $IrMn_3$ layers contribute to the NMSHE-driven spin accumulation at the $IrMn_3$/Pt interface.

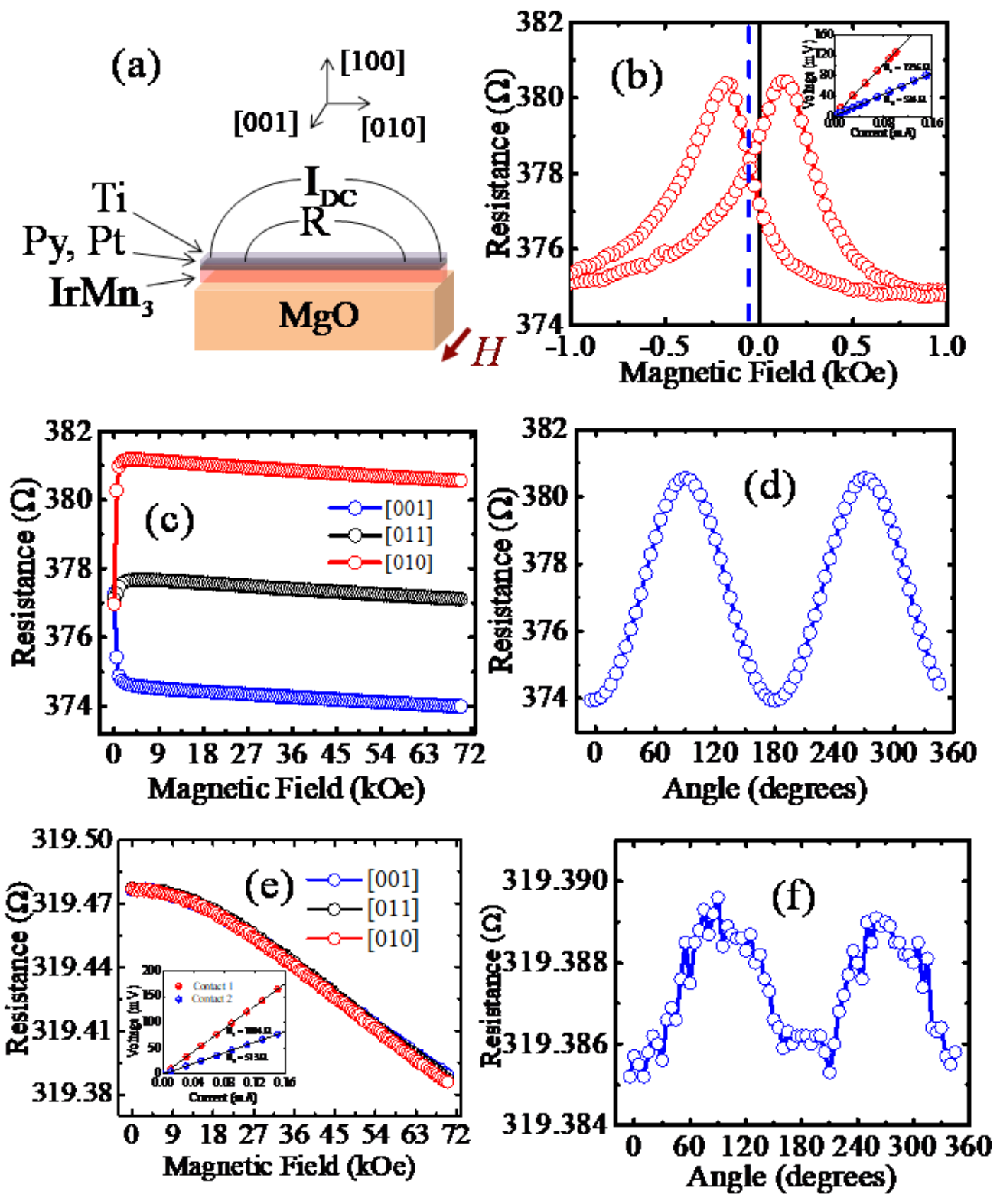

**Figure 3:** Transport measurements performed at a temperature of 10 K with a DC current of 0.1 mA. **(a)** Schematic of the experimental configuration used for the transport measurements, indicating the crystallographic directions of the antiferromagnetic $IrMn_3$ layer; **(b)** Resistance as a function of the applied magnetic field, showing a shift associated with the exchange bias field. The inset confirms that both electrical contacts are ohmic; **(c)** and **(e)** Resistance as a function of magnetic field in the range 0–70 kOe measured along three different crystallographic directions for the $IrMn_3$/Py and $IrMn_3$/Pt samples, respectively. The inset in (e) confirms ohmic behavior of the contacts; **(d)** and **(f)** Angular dependence of the resistance measured under an applied magnetic field of 70 kOe for the $IrMn_3$/Py and $IrMn_3$/Pt samples, respectively.

The temperature dependence of the exchange bias field was further investigated using transport measurements. In **Fig. 4**, a DC current of 0.05 mA was applied to analyze the influence of the magnetic spin Hall effect (MSHE) on the anisotropic magnetoresistance (AMR) at T=75 K, close to the surface magnetic transition discussed in **Fig. 2**. Additional measurements were performed at T=300 K, a temperature at which the $IrMn_3$ surface moments are expected to be highly unstable, resulting in reduced exchange energy. **Figs. 4(a)** and **4(b)** present transport measurements performed after a 70 kOe field-cooling process, showing the temperature dependence of the coercivity and exchange bias field, respectively. These results are consistent with the hysteresis-loop measurements shown in **Fig. 2(c)**. The insets of **Figs. 4(a)** and **4(b)** display the derivatives of the resistance with respect to the magnetic field, indicating the field-sweep direction at T=75 K and T=300 K, respectively. The exchange coupling at the antiferromagnetic interface with the adjacent material, as well as the resulting antiferromagnetic configurations, plays a crucial role in the manipulation of spin transport at the interface [44, 45]. In **Figs. 4(c)**, **4(d)**, **4(e)**, and **4(f)**, no pronounced intrinsec oscillations are observed. However, at T=300 K a weak non-linear behavior appears (indicated by the orange arrow in **Fig. 4(e)**), suggesting that the MSHE is approaching the threshold required to induce oscillations in the AMR. A similar behavior is observed in the angular dependence of the resistance measured at T=75 K for applied magnetic fields of 30, 70, and 85 kOe, as shown in **Figs. 4(c)** and **4(d)**, where the resistance variation remains smooth. In contrast, at T=300 K, distinct shifts in the resistance amplitude as a function of the applied field angle and magnetic field strength are observed (**Fig. 4(f)**). This behavior provides further evidence that the AMR approaches an intrinsec oscillatory regime at high temperatures or DC current.

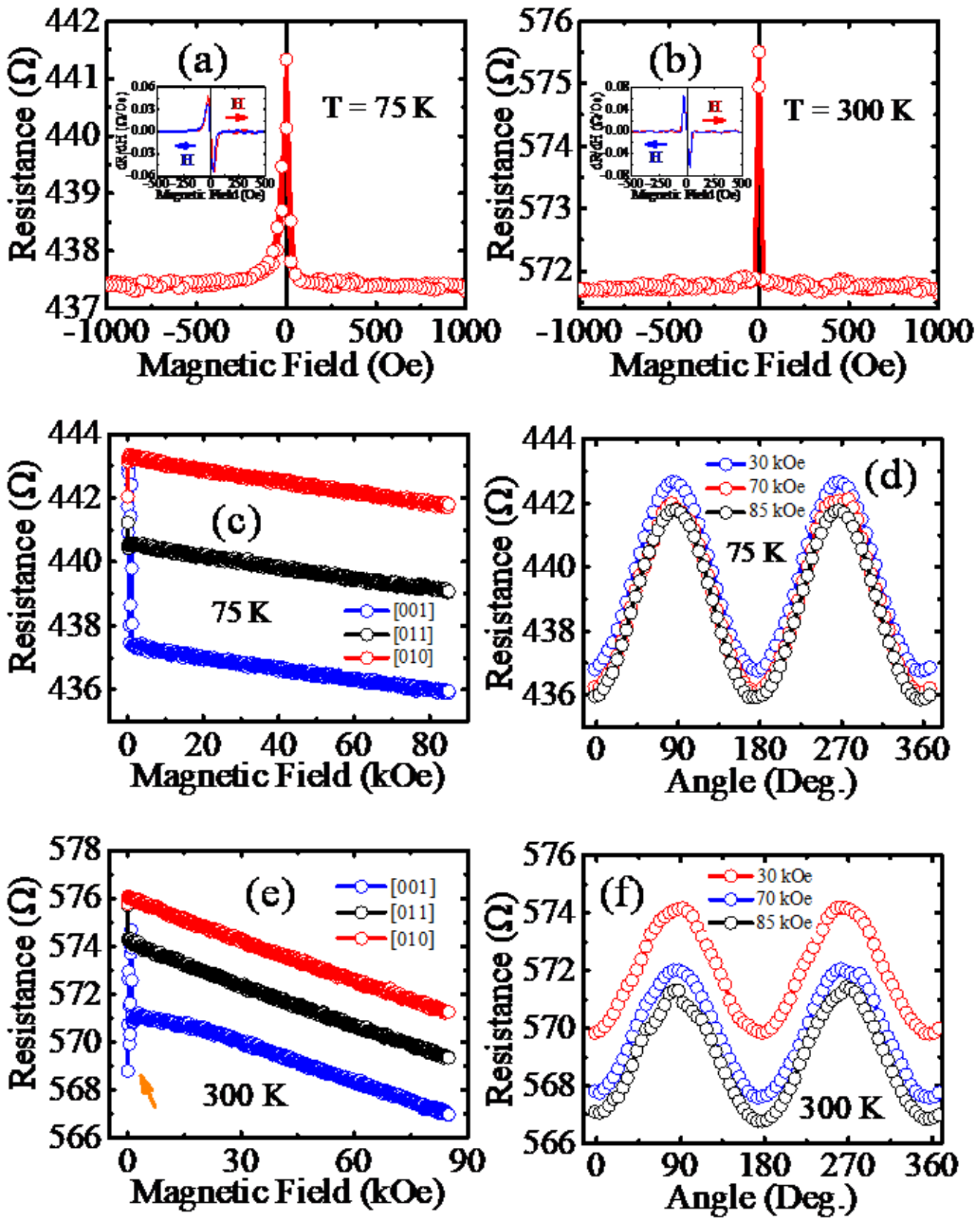

**Figure 4:** Measurements were performed on a typical sample with the structure (100)MgO/$IrMn_3$(10 nm)/Py(6 nm)/Ti(2 nm). **(a)** and **(b)** show the resistance as a function of the applied magnetic field at temperatures of 75 K and 300 K, respectively. The insets highlight the small magnetic-field shift associated with the exchange bias. The red and blue curves indicate the magnetic-field sweep directions; **(c)** and **(e)** present the resistance as a function of magnetic field in the range 0–8.5 kOe measured along three different crystallographic directions at temperatures of 75 K and 300 K, respectively; **(d)** and **(f)** display the angular dependence of the resistance measured under applied magnetic fields of 30, 70, and 85 kOe at temperatures of 75 K and 300 K, respectively.

Electrical manipulation in antiferromagnetic spintronics [46, 47] has demonstrated that interfacial effects are of fundamental importance for spintronic device applications. One such promising application is the realization of oscillations at antiferromagnet/adjacent-material interfaces. In general, oscillation phenomena play a central role in spintronics, as they underpin mechanisms for information generation, transmission, and processing. For these reasons, the exploration of new methods to observe and control oscillations is of significant interest to the scientific community. **Fig. 5** illustrates the general experimental configuration used for the oscillation measurements presented in this work, representing a novel approach to observing such phenomena. The oscillations detected here through anisotropic magnetoresistance (AMR) measurements in $IrMn_3$/adjacent-material heterostructures originate from uncompensated Mn spins at the surface of the $IrMn_3$ layer. In this class of non-collinear antiferromagnets, the antiferromagnetic order at the surface can differ substantially from that of the bulk and may lose its non-collinear character depending on temperature. Consequently, the surface magnetic order enables the emergence of additional phenomena, such as oscillations, that are not present in the bulk antiferromagnetic state. As schematically shown in **Fig. 5**, the magnetic order of the $IrMn_3$ film interior differs from that of the surface atoms. The deposited adjacent material (Py or Pt) therefore couples primarily to the surface magnetic moments of $IrMn_3$, giving rise to two distinct contributions: one associated with the surface uncompensated spins and another originating from the bulk antiferromagnetic order of $IrMn_3$.

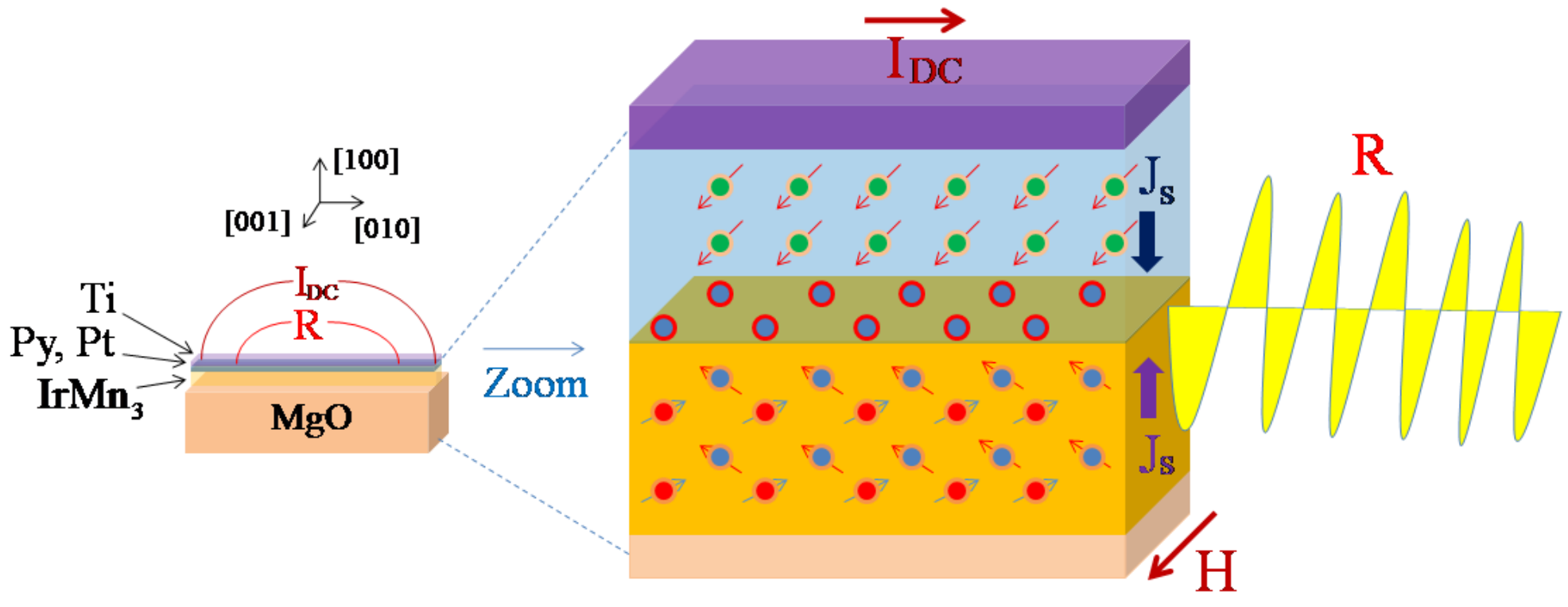

**Figure 5:** Schematic of the experimental setup employed to investigate the oscillation effect. The resulting spin-current flux is determined by the nature of the material adjacent to $IrMn_3$.

As is well known, a spin-polarized charge current can generate different effects depending on the material system. When permalloy is used as the material adjacent to $IrMn_3$, two contributions are responsible for the observation of oscillations: the anisotropic magnetoresistance (AMR) of permalloy and the magnetic spin Hall effect (MSHE). These effects act at the $IrMn_3$/Py interface, where the MSHE produces a spin accumulation that exerts spin torques on the permalloy magnetization. These torques can induce oscillations in the interfacial magnetization, which can be detected through AMR measurements, since the interfacial magnetization is directly reflected in the resistance signal. In contrast, when platinum is used as the adjacent material, the observation of auto-oscillations arises from different contributions: an AMR generated by interfacial effects, the non-magnetic spin Hall effect (NMSHE) of $IrMn_3$, and the NMSHE of platinum. In this case, spin accumulation occurs on both sides of the $IrMn_3$/Pt interface. The oscillation behavior is presented in **Fig. 6**, with measurements performed using a DC current of 0.1 mA at a temperature of 300 K. The oscillations are found to be independent of the crystallographic directions ([001], [011], and [010]), indicating that the underlying mechanisms follow the polarization of the applied magnetic field. This independence further implies that magnetocrystalline anisotropy does not play a dominant role in polarizing the spin current in this regime.

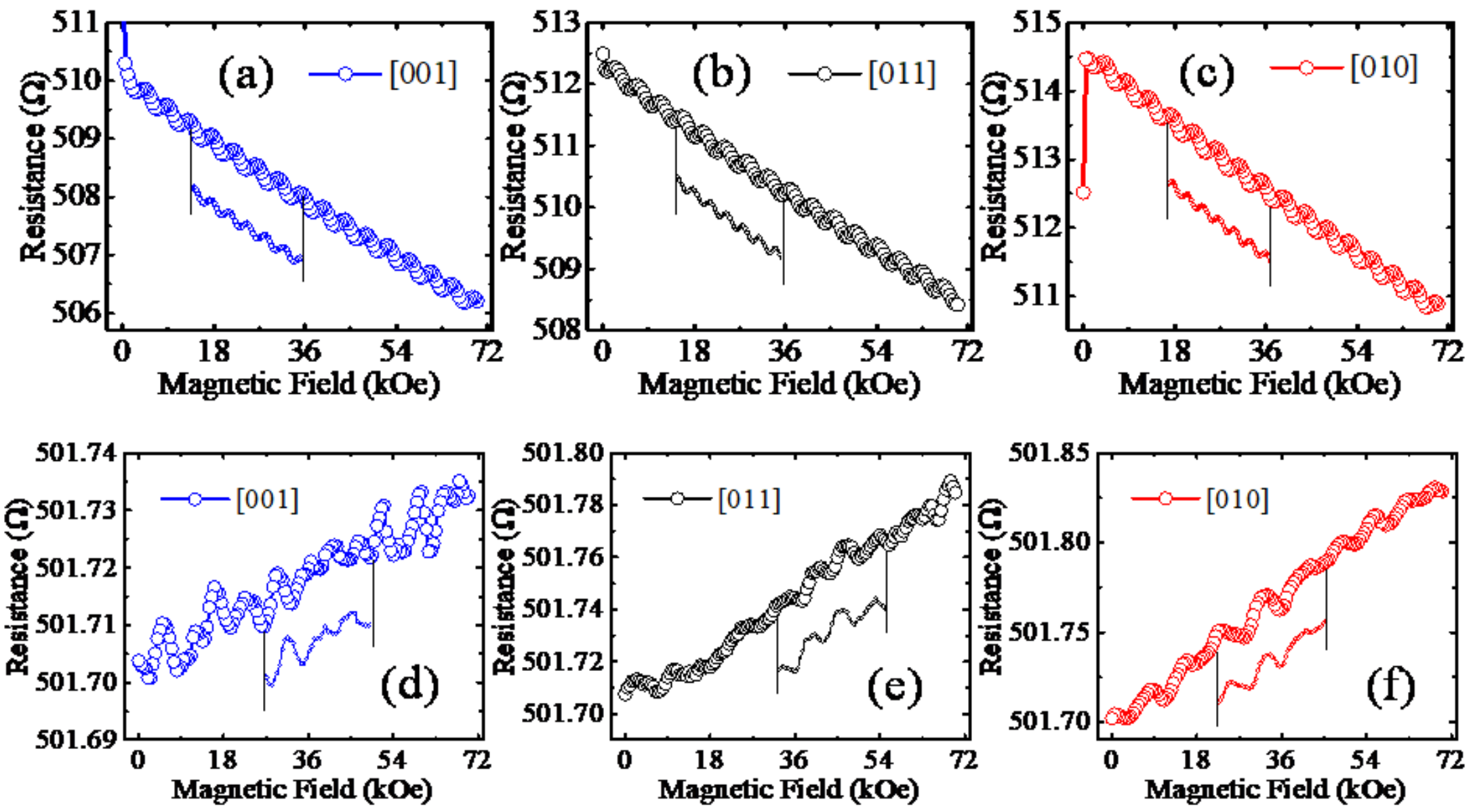

**Figure 6:** Transport measurements performed at 300 K under a DC current of 0.1 mA are shown for different crystallographic directions. Panels **(a)–(c)** present data for a typical (100) $MgO/IrMn_3$(10 nm)/Py(6 nm)/Ti(2 nm) sample, whereas panels **(d)–(f)** correspond to a (100) $MgO/IrMn_3$(10 nm)/Pt(6 nm)/Ti(2 nm) sample.

**Figs. 6(a)**, **6(b)**, and **6(c)** present measurements performed on a representative (100) MgO/$IrMn_3$(10 nm)/Py(6 nm)/Ti(2 nm) sample. In these data, the oscillation behavior is well defined and exhibits regular periodicity. This behavior can be attributed to the ferromagnetic nature of permalloy, which allows the oscillations detected via AMR to occur with well-defined periods. In contrast, **Figs. 6(d)**, **6(e)**, and **6(f)** show measurements obtained from a representative (100) MgO/$IrMn_3$(10 nm)/Pt(6 nm)/Ti(2 nm) sample. In this case, the oscillations do not display well-defined periodic behavior. This difference arises from the significantly lower AMR amplitude, as the signal is generated primarily by interfacial effects involving $IrMn_3$ and Pt atoms rather than by an ferromagnetic contribution. As discussed in **Fig. 4**, the oscillatory behavior depends on the applied DC current and, consequently, on the strength of the MSHE and NMSHE. These results demonstrate that both the magnetic and non-magnetic spin Hall effects provide effective means to control the AMR response, and thus the oscillation behavior, in $IrMn_3$/adjacent-material heterostructures. The present measurements indicate a current- and temperature-dependent onset of the oscillatory regime; however, the available current sweep and the present device geometry do not allow us to determine a quantitative critical current density (Jc). A systematic current-dependent study would be required to establish the threshold current and the scaling of oscillation frequency and amplitude with $I_{DC}$. Furthermore, we state that the oscillations to which we refer are a spontaneous periodic or quasi-periodic variation in the measured electrical resistance that develops under direct current (DC) electrical polarization, without the application of an externally imposed microwave magnetic field or alternating current (AC) magnetic excitation.

## Conclusion

Spintronics is a field of physics in which the transfer, encoding, and processing of information are governed not by the electron charge, but by its spin degree of freedom. Spintronic devices typically consist of materials with distinct magnetic and electrical properties, where reduced dimensionality plays a crucial role. Among the many active research topics in spintronics, oscillations represent a particularly important class of dynamical phenomena. Conventionally, such oscillations are achieved through precise control of damping in ferromagnetic or ferrimagnetic materials. In this work, we demonstrate a new method for observing oscillations based on magnetoresistance measurements. The effect emerges at the interface between a non-collinear antiferromagnet and an adjacent ferromagnetic or paramagnetic material. When permalloy is used as the adjacent layer, the oscillating magnetoresistance is primarily driven by the magnetic spin Hall effect (MSHE), whereas in the case of platinum, the non-magnetic spin Hall effect (NMSHE) dominates. The intensity of the oscillations is found to scale with the applied DC current. The physical origin of these oscillations lies in spin accumulation at the antiferromagnetic/ferromagnetic or antiferromagnetic/paramagnetic interface. Depending on the nature of the adjacent material, spin torques act at the interface either unidirectionally or bidirectionally, leading to sustained oscillations of the interfacial magnetization. These spin-torque-driven dynamics manifest experimentally as oscillations in the anisotropic magnetoresistance.

**Acknowledgements**

This research was supported by Conselho Nacional de Desenvolvimento Científico e Tecnológico (CNPq) with Grant Number: 300631/2025-1, Coordenação de Aperfeiçoamento de Pessoal de Nível Superior (CAPES) with Grant Number: PROAP2025UFRPE, and Fundação de Amparo à Ciência e Tecnologia do Estado de Pernambuco (FACEPE) with Grant Number: APQ-1397-3.04/24.

**Contributions**

M. C., J. L., C. E., and K. M. analyzed all the experimental measures and J. H. discussed, wrote and supervised the work.

**Conflict of interest**

The authors declare that they have no conflict of interest.

**Data Availability Statement**

Data will be made available on reasonable request.